\documentclass{ws-p8-50x6-00}
\newcommand{\elel}{\ensuremath{\mathrm{e^+e^-}}}
\newcommand{\gaga}{\mathrm{\gamma\gamma}}
\newcommand{\qq}{\mathrm{q\bar{q}}}
\newcommand{\eeqq}{\mathrm{e^+e^-\rightarrow q\bar{q}}}

\newcommand{\Rl}{\mathrm{R_l}}
\newcommand{\Rel}{\mathrm{R_el}}
\newcommand{\Rmu}{\mathrm{R_\mu}}
\newcommand{\Rtau}{\mathrm{R_\tau}}
\newcommand{\Afbl}{\mathrm{A_{FB}^{0,l}}}
\newcommand{\Afbe}{\mathrm{A_{FB}^{0,e}}}
\newcommand{\Afbmu}{\mathrm{A_{FB}^{0,\mu}}}
\newcommand{\Afbtau}{\mathrm{A_{FB}^{0,\tau}}}
\newcommand{\MZ}{\mathrm{M_Z}}
\newcommand{\MH}{\mathrm{M_H}}
\newcommand{\GZ}{\mathrm{\Gamma_Z}}
\newcommand{\sigh}{\mathrm{\sigma^0_{had}}}
\newcommand{\sigl}{\mathrm{\sigma^0_{l}}}
\newcommand{\lept}{\mathrm{l^+l^-}}
\newcommand{\mumu}{\mathrm{\mu^+\mu^-}}
\newcommand{\tautau}{\mathrm{\tau^+\tau^-}}
\newcommand{\Aleph} {{\sc Aleph} }%
\begin{document}

\title{Electroweak results from the Z resonance Cross-Sections and Leptonic
Forward-Backward Asymmetries with the ALEPH detector}

\author{E. Tournefier}

\address{CERN, Switzerland
\\E-mail: Edwige.Tournefier@cern.ch}

\maketitle

\abstracts{The measurement of the Z resonance parameters and lepton 
forward-backward asymmetries are presented. These are determined
from a sample of 4.5 million Z decays accumulated with the \Aleph detector
at LEP~I.}

\section{Introduction}
\label{intro}
From 1990 to 1995 the LEP $\elel$ storage ring was operated at centre of 
mass energies close to the Z mass, in the range $|\sqrt{s}-\MZ | <3$~GeV.
Most of the data have been recorded at the maximum of the resonance 
(120~pb$^{-1}$ per experiment) and about 2~GeV below and above (40~pb$^{-1}$ 
per experiment).

The measurement of the hadronic and leptonic cross sections as well as the  
leptonic forward backward asymmetries performed with the \Aleph detector
at these energies are presented here.  
The large statistic allows  a precise measurement of these quantities
which are then used to determine the Z lineshape parameters: the Z mass $\MZ$, 
the Z width $\GZ$, the total hadronic cross section at the pole $\sigh$ and the 
ratio of hadronic to leptonic pole cross sections $\Rl=\sigh/\sigl$.

Here we will give some details of the \Aleph experimental measurement 
of these quantities. A review of the whole LEP electroweak measurements and 
a discussion of the results as a test of the Standard Model can be found
in another talk of this conference~\cite{Pinfold}.
%
%\section{Luminosity}
%\label{lumi}
%
\section{Cross sections and leptonic Forward-Backward asymmetries measurement}
\label{xs_afb}
The cross section and asymmetries are determined for  the s-channel process
$\mathrm{e^+e^- \rightarrow Z,\gamma \rightarrow f\bar{f}}$.
The cross section  is derived from the number of selected events
$\mathrm{N_{sel}}$ with
\begin{equation}
\mathrm{ \sigma_{f\bar{f}} = \frac{N_{sel}(1-f_{bkg})}{\epsilon}
\frac{1}{\cal L}}
\end{equation}
where $\mathrm{f_{bkg}}$ is the fraction of background, 
$\mathrm{\epsilon}$ is the selection efficiency and ${\cal L}$ is the integrated
luminosity.
Note that in the case of $\elel$ final state the irreducible background originating
from the exchange of $\gamma$ (Z) in the t-channel is subtracted from $\mathrm{N_{sel}}$
to obtain the s-channel cross section.

The leptonic forward-backward asymmetry ($\mathrm{A_{FB}}$) is derived 
from a fit to the angular distribution
\begin{equation}
\mathrm{ \frac{d\sigma_{f\bar{f}}}{d\cos\theta^*}\propto 
1+cos^2\theta^*+\frac{8}{3}A_{FB}cos\theta^*}
\end{equation}
where $\theta^*$ is the centre of mass scattering angle between the incoming
$\mathrm{e^-}$ and the out-coming negative lepton.
 
To achieve a good precision on the cross section a high efficiency  and low 
background is necessary while the asymmetry is insensitive to the overall
efficiency and to a background with the same asymmetry as the signal.
This justifies the use of different leptonic selections for cross section 
and asymmetry measurement.
\subsection{Hadronic cross sections}
\label{xs_qq}
Four million of $\eeqq$ events have been recorded at the Z peak, 
leading to a statistical  uncertainty of $0.05\%$. 
The systematic uncertainty has been reduced to the same level.

The hadronic cross section measurement is based on two independent
selections. 
The first selection is based on charged track properties
while the second is based on calorimetric energy. 
Details of these selections can be found in previous publications~\cite{AlephLS}.
These two measurements are in good agreement and are combined 
to obtain the final result.
The systematic uncertainties 
of these selections are almost uncorrelated because they are mainly based on 
uncorrelated quantities therefore 
 the combination of the 2  selections allows to reduce 
the systematic uncertainty.

\begin{figure}[htpb]
\vspace{-.6cm}
\begin{center}
\epsfxsize=15pc
\vspace{-1cm}
\epsfbox{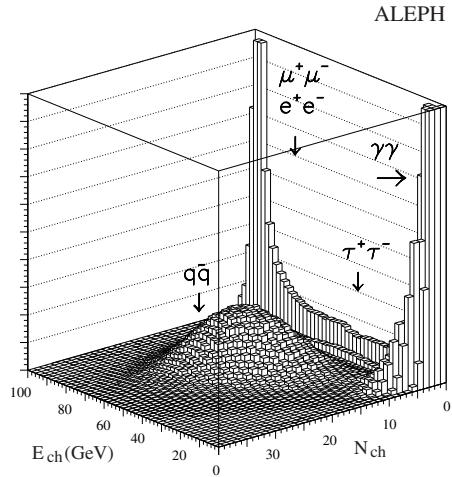}
\end{center}
\caption{Distribution of charged track multiplicity ($\mathrm{N_{ch}}$)
versus charged track energy ($\mathrm{E_{ch}}$). \label{TPChad}}
\end{figure}
Figure~\ref{TPChad} shows the distribution of charged multiplicity versus
the charged track energy for signal and background. These variables are used
to separate $\eeqq$ events from background in the charged track selection.
The most dangerous background in both selections is $\gaga$ events because 
Monte Carlo prediction is not fully reliable. Therefore a method to determine
this background from the data has been developed. This is achieved by 
exploiting the different $\sqrt{s}$ dependence of the resonant (signal) 
and non-resonant (background) contributions. The resultant
systematic error ($0.04\%$) reflects the statistic of the data.

%
%\begin{figure}[t]
%\begin{center}
%\epsfxsize=15pc
%\vspace{-1cm}
%\epsfbox{etpc-ntpc.eps}
%\end{center}
%\caption{Distribution of charged track multiplicity ($\mathrm{N_{ch}}$)
%versus charged track energy ($\mathrm{E_{ch}}$). \label{TPChad}}
%\end{figure}
%
%
\begin{table}[h]
\vspace{-.4cm}
\caption{Efficiency, background and systematic errors for the two hadronic selections at the peak point.}
\begin{center}
\begin{tabular}{|c|c|c|}
\hline
               & { Charged tracks} & { Calorimeter}\\
\hline
{ Efficiency ($\%$) }   & $97.48$     &  $99.07$ \\
\hline
{ Background ($\%$)}& & \\
$\tau^+\tau^-$  & 0.32 & 0.44 \\
$\gamma\gamma$  & 0.26 & 0.16 \\
$e^+e^-$        &   -  & 0.08 \\ 
\hline
{ Systematic ($\%$)   } & &\\
Detector simulation  & 0.02 & 0.09\\
Hadronisation modelling  & 0.06 & 0.03\\
$\tau^+\tau^-$ bkg & 0.03 & 0.05 \\
$\gamma\gamma$ bkg & 0.04 & 0.03 \\
$e^+e^-$       bkg &   -  & 0.03 \\ 
\hline
Total syst. & 0.087 & 0.116\\
{ Combined  }  &
 \multicolumn{2}{c|}{ 0.071}\\
\hline
 \multicolumn{3}{c}{ }\\
 \multicolumn{3}{c}{ }\\
\end{tabular}
\end{center}
\label{qqsys}
\end{table}
The dominant systematic error in the calorimetric selection comes from the 
calibration of calorimeters ($0.09\%$)
and, in the charged
track selection from the determination of the acceptance ($0.06\%$).
Table~\ref{qqsys} gives a breakdown of the efficiency, the background and 
the systematic uncertainties of both selections.
\subsection{Leptonic cross sections}
\label{xs_ll}
The statistical uncertainty in the leptonic channel is of the order of $0.15\%$,
The aim of the analysis is to reduce the systematic to less than $0.1\%$.
Two analyses were developed for the measurement of the leptonic cross sections.
The first one, referred to as {\it exclusive} is based on three independent
selections each aimed at isolating one lepton flavour and still follows the general
philosophy of  the analysis procedures described earlier~\cite{AlephLS}.
The second one is new and 
has been optimised for the measurement of $\Rl$, it is refered 
as {\it global} di-lepton selection.
The results of these selections agree within the uncorrelated statistical error
and have been combined for the final result.
These selections are not independent since they make use of 
 similar variables therefore their combination does not  
reduce the systematic uncertainty. 

We concentrate here on the {\it global} analysis.
This selection takes advantage of the excellent particle identification capabilities
(dE/dx, shower developement in the calorimeters and muon chamber information) 
and the high granularity of the \Aleph detector.
First di-leptons are selected within the detector acceptance with an efficiency
of $99.2\%$ and the background arising from $\gaga$, $\qq$ and cosmic events
is reduced to the level of $0.2\%$. Then the lepton flavour separation is 
performed inside the di-lepton sample so that the systematic uncertainties 
are anti-correlated between 2 lepton species and that no additional uncertainty
is introduced on $\Rl$.
Table~\ref{globalsys} gives a breakdown of the systematic errors obtained with 1994 data.
The background from $\qq$ and $\gaga$ events affects mainly the $\tautau$ channel,
therefore this channel is affected by bigger selection systematics 
than $\elel$ and $\mumu$. 

As an  example we consider the systematic errors related to 
the $\tautau$ selection efficiency.  This efficiency is measured on the data:
$\tautau$ events are selected using tight selection criteria to flag $\tau$-like
hemispheres; with the sample of opposite hemispheres, {\it artificial} $\tautau$
events are constructed by associating two back-to-back such hemispheres and
the selection cuts  are applied. In order to assess the validity of the
method and to correct for possible bias of this method, two different 
Monte Carlo reference samples are used. On the first sample the same procedure of 
artificial $\tautau$ events is applied, on the second one the selection cuts 
are applied directly. The uncertainty on the efficiency measured with this method
is dominated by the statistic of the artificial events used in the data.
This method is applied in order to measure the inefficiency arising from
$\qq$ cuts and in the flavour separation.

The dominant systematic in $\elel$ channel arises from t-channel subtraction
and is given by the theoretical uncertainty~\cite{alistar} 
on the t-channel contribution to the cross section.

This leptonic cross section measurement contributes to a systematic uncertainty 
on $\Rl$ of $0.08\%$.
%\renewcommand{\arraystretch}{1.35}
%\setlength{\tabcolsep}{2mm}
%\begin{table}[t]
\begin{table}[t]
%\footnotesize
\caption{ Systematic uncertainties in percent of dilepton cross sections for 
peak 1994 data. Correlations between lepton flavours are taken into account
in the $\lept$ column. \label{globalsys}}
\begin{center}
%\vspace{-1.cm}
%\mbox{
\begin{tabular}{|l|c|c|c||c|}
\hline
            & $\elel$ & $\mumu$ & $\tautau$ & $\lept$ \\
\hline
\multicolumn{5}{|c|}{ Global selection}\\
\hline
Tracking efficiency&    0.05  & 0.03   & 0.03&0.04 \\
\hline
Angles measurement (*)&0.02 & 0.01 & 0.01 & 0.02 \\
\hline
ISR and FSR simulation (*)& 0.03 & 0.03 &0.03 & 0.03 \\
\hline
$\gaga$ cuts (*)& 0.02 & - & 0.05 & 0.02 \\
\hline
$\qq$ cuts& - & - & 0.11  & 0.04 \\
\hline
$\gaga$ background (*) & - & - & 0.02 & -\\
\hline
$\qq$ background(*) & - & - & 0.04 & 0.01 \\
\hline
\multicolumn{5}{|c|}{Flavour separation}\\
\hline
$\mumu$/$\tautau$ & - & 0.03 & 0.03 & - \\
\hline
$\elel$/$\tautau$& \multicolumn{4}{|c|}{ }\\
$cos\theta^*<0.7$& 0.08 & - & 0.07 & 0.01\\
$cos\theta^*\geq0.7$& - & - & 0.06 & 0.02\\
\hline
\multicolumn{5}{|c|}{t-channel subtraction}\\
\hline
(*) & 0.11&-&-& 0.04 \\
\hline
\multicolumn{5}{|c|}{Monte Carlo statistic}\\
\hline
 & 0.05 & 0.06 & 0.07 & 0.04 \\
\hline
\hline
Total & 0.16 & 0.08 &0.19 & 0.09 \\
\hline
\multicolumn{5}{c}{(*)\footnotesize uncertainties completely correlated among all energy points.}
\end{tabular}
%}
\end{center}
\end{table}

\subsection{Leptonic Forward-Backward asymmetries}
\label{Afb}
The measurement of the asymmetries is dominated by the statistical
uncertainty equal to 0.0015.
Special muon and tau selections have been 
designed for the asymmetry measurement while the $\elel$ {\it exclusive}
selection is used for the $\elel$ asymmetry  measurement.
 The $\elel$ angular distribution needs to be corrected for efficiency
before subtracting the t-channel and therefore relies on Monte Carlo
while $\mumu$ and $\tautau$ angular distributions do not need to be corrected
with Monte Carlo since the selections are designed so that the 
efficiency is symmetric.
 
Because of the $\gamma-Z$ interference, 
 the asymmetry varies rapidly with the centre of mass energy
$\sqrt{s'}$ around the Z mass. Cuts on energy induce a dependence of the 
efficiency with $\sqrt{s'}$ and therefore  could introduce a bias in the measurement
since the efficiency would no more be symmetric. To minimise these effects
the selections are mainly  based on particle identification instead of 
kinematic variables. 
\begin{figure}[htpb]
\begin{center}
\epsfxsize=17pc
%\vspace{-1cm}
\epsfbox{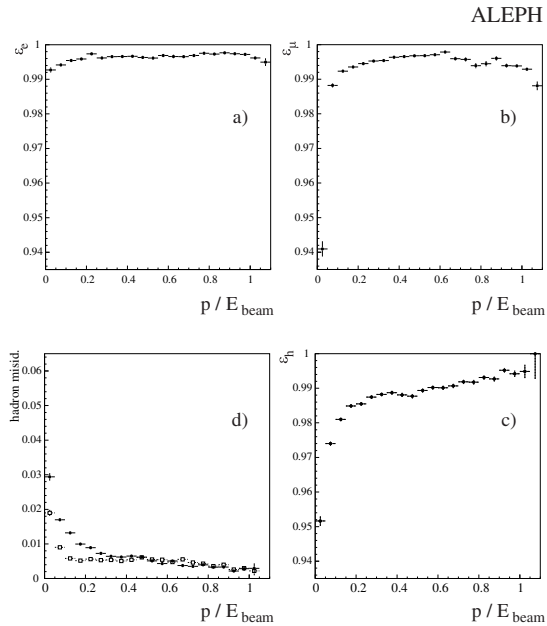}
\end{center}
\caption{Identification efficiencies for a) electrons, b) muons and c) pions
as a function of the momentum to the beam energy ratio. In d) the probability for
a pion to be misidentified as an electron (squares) or a muon (circles) is shown.
\label{leptid}}
\end{figure}
Figure \ref{leptid} shows the particle identification efficiency as a function 
of energy.

The asymmetry is extracted by performing a maximum
likelihood fit to the differential cross section.
The dominant systematic uncertainty arises from t-channel subtraction in the 
Bhabha channel (0.0013 on $\Afbe$) and other systematic errors are smaller than 0.0005.
\begin{figure}[htpb]
\vspace{-.8cm}
\hspace{-2cm}
\begin{minipage}[b]{\textheight}
\begin{center}
\epsfxsize=19pc
\epsfbox{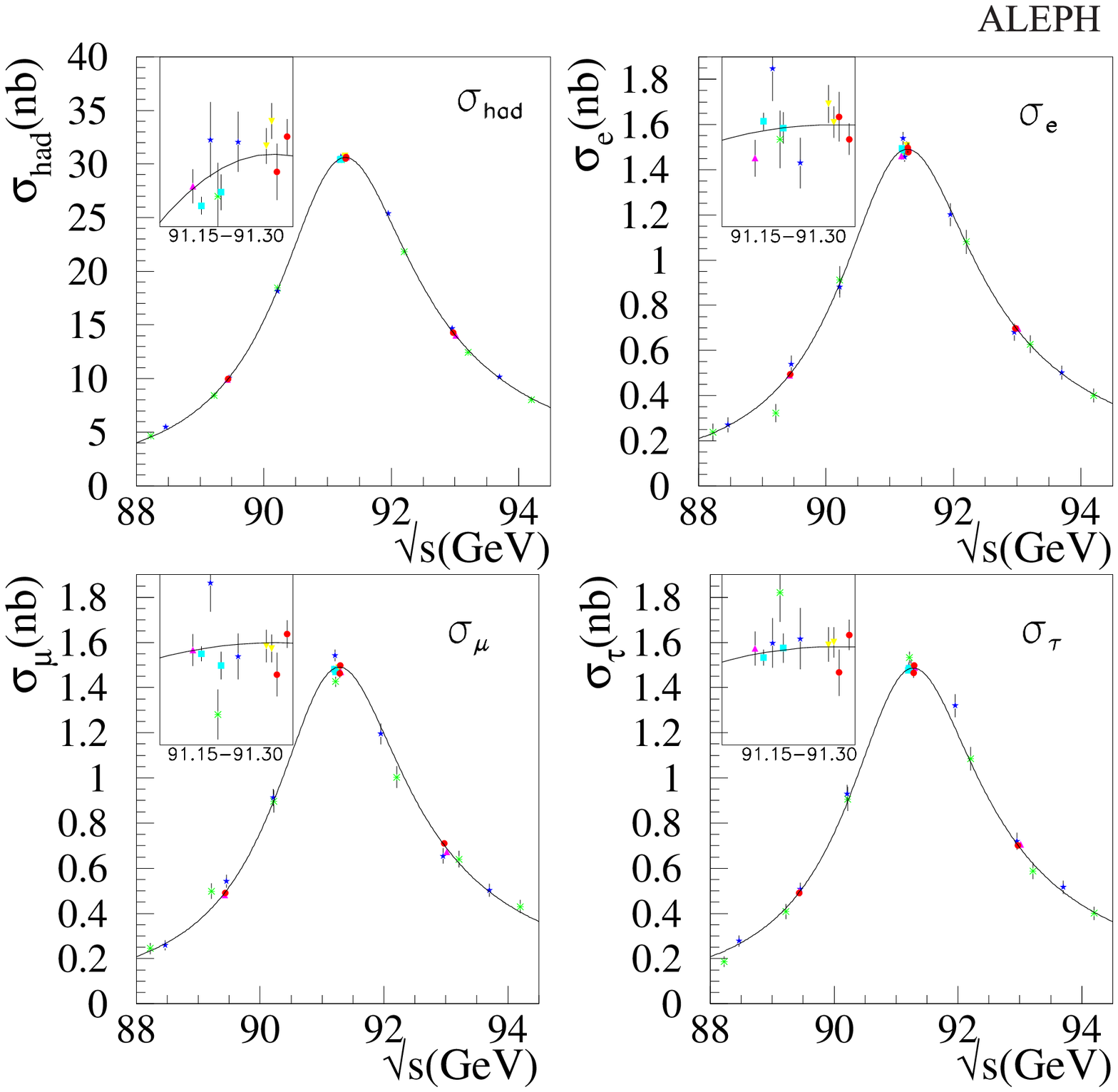}
\end{center}
\caption{Measurement of cross sections. The inserts show enlarged views of the peak
region. \label{lineshape}}
\end{minipage}
%\end{figure}
%
%\begin{figure}[t]
\vspace{-1cm}
\hspace{-2cm}
\begin{minipage}[b]{\textheight}
\begin{center}
\epsfxsize=19pc
\epsfbox{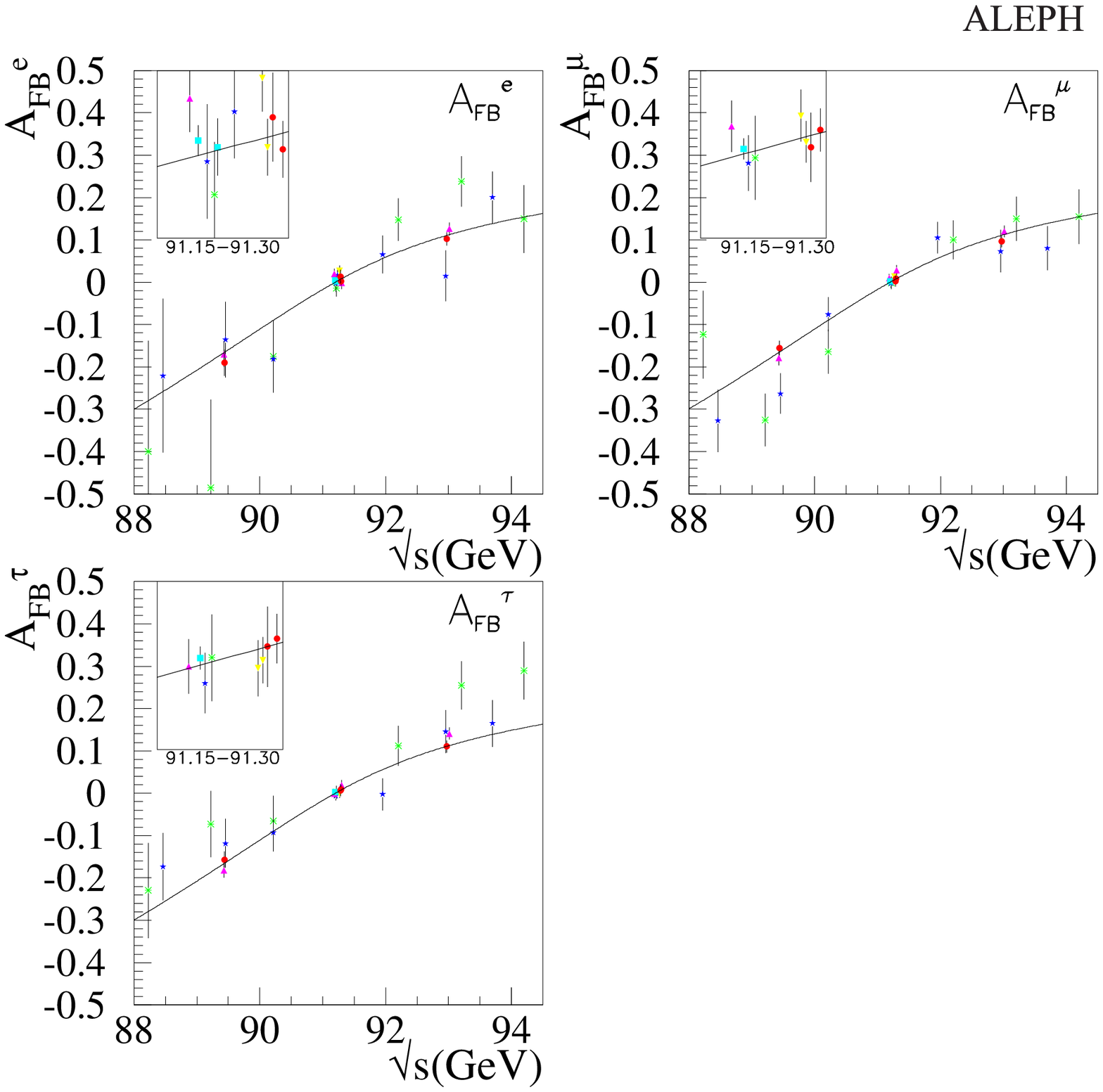}
\end{center}
\caption{Measurement of the asymmetries.
The inserts show enlarged views of the peak region. \label{asymmetries}}
\end{minipage}
\end{figure}
\section{Results}
\label{results}
Figures~\ref{lineshape} and \ref{asymmetries} show the measured cross sections 
and asymmetries.
The Z lineshape parameters are fitted to these measurements with the latest version of
 ZFITTER~\cite{ZFITTER}. 
The error matrix used in the $\chi^2$ fit includes
 the experimental statistic and systematic uncertainties,
the LEP beam energy measurement  uncertainty and
the theoretical uncertainties arising from the small angle Bhabha
cross section in the luminosity determination
and the t-channel contribution to wide angle Bhabha events.
The results are shown in Table~\ref{res_tab}.

\begin{table}[t]
\caption{Result of the fit to experimental measurement of cross sections and leptonic
forward-backward asymmetries. The total error is splitted  into statistical and 
systematic errors, theoretical error on luminosity and on t-channel
and LEP beam energy measurement error.\label{res_tab}}
\begin{center}
\begin{tabular}{|c|c|r|r|r|r|r|}
\hline
                 & value             & stat   &   exp  & { lumi}  & { t-ch} & { $E_{beam}$} \\ 
\hline
$\MZ$           & 91.1883$\pm$0.0031 & 0.0024  & 0.0002 &       &       & { 0.0017}  \\
\hline
$\GZ$           & 2.4953$\pm$0.0043 & 0.0038 & 0.0009 &       &       &{ 0.0013}\\
\hline
$\sigh$        & 41.557$\pm$0.058  & 0.030  & 0.026  & {0.025} &       &{ 0.011}\\
\hline
$\Rel$           & 20.677$\pm$0.075  & 0.062  & 0.033  &       & { 0.025} & { 0.013}\\
\hline
 $\Rmu$      & 20.802$\pm$0.056  & 0.053  & 0.021  &       &       & {0.006}\\
\hline
 $\Rtau$     & 20.710$\pm$0.062  & 0.054  & 0.033  &       &       & {0.006}\\
\hline
 $\Afbe$        & 0.0189$\pm$0.0034 & 0.0031 & 0.0006 &       & { 0.0013}& { 0.0002}\\
\hline
$\Afbmu$        & 0.0171$\pm$0.0024 & 0.0024 & 0.0005 &       &       & { 0.0002}\\
\hline
$\Afbtau$       & 0.0169$\pm$0.0028 & 0.0026 & 0.0011 &       &       & { 0.0002}\\
\hline
\hline
$\Rl$           & 20.728$\pm$0.039 & 0.033  & 0.020  &       & { 0.005} & { 0.002}\\
\hline
$\Afbl$          & 0.0173$\pm$0.0016 & 0.0015  & 0.0004  &       & { 0.0002} & { 0.0001}\\
\hline
\end{tabular}
\end{center}
\end{table}

\begin{figure}[t]
\begin{center}
\epsfxsize=15pc
\epsfbox{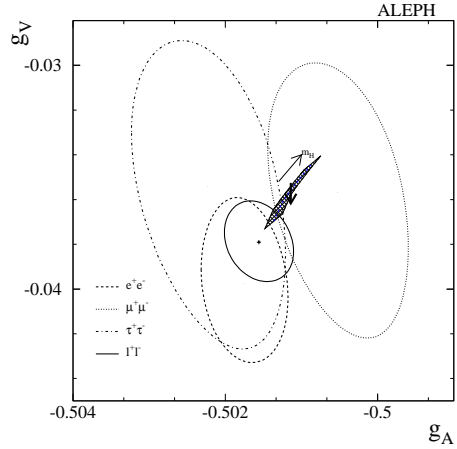}
\end{center}
\caption{Effective lepton couplings. Shown are the one-$\sigma$ contours. The shaded
area indicates the Standard Model expectation for $\mathrm{M_t}=174\pm5$~GeV/$c^2$
and $90< \MH GeV/c^2<1000$; the vertical fat arrow shows the change if the 
electromagnetic coupling constant is varied within its error.
The sign of the couplings were assigned in agreement with $\tau$ polarisation measurements
and neutrino data scattering.
\label{gvga}}
\end{figure}

The value of the Z couplings to charged leptons
$|g_V|$ and $|g_A|$ can be derived from these parameters. 
The experimental measurement is shown in Figure \ref{gvga}
with the Standard Model prediction. The data favor a light Higgs.

The value of $\alpha_s$ can also be extracted from $\Rl$, $\GZ$ and 
$\sigh$:
\begin{equation}
\mathrm{\alpha_s = 0.115 \pm 0.004_{exp} \pm 0.002_{QCD} }
\end{equation}
where the first error is experimental and the second reflects uncertainties
on the QCD part of the theoretical prediction~\cite{ALPHAS}. 
Here the Higgs mass has been fixed to 150 GeV and the dependence of $\alpha_s$
with $\MH$ can be approximately parametrised by 
$\mathrm{\alpha_s(M_H)=\alpha_s(M_H=150 GeV)\times(1+0.02\times ln(M_H/150))}$
where $\MH$ is expressed in GeV.
\section{Conclusion}
\label{conclusion}
The high statistic accumulated by \Aleph during LEP 1 running allows to measure the 
hadronic and leptonic cross sections and the leptonic forward-backward asymmetries
with statistical and systematic precision of the order of 1 permil.
These measurements are turned into precise determination of the Z boson
properties and constraints on the Standard Model parameters.
\section*{Acknowledgments}
I would like to thank D. Schlatter for his help in preparing this talk.
I also would like to thank the organisers of the Lake Louise Winter Institute
conference for the interesting program and the nice atmosphere at the conference.

\end{document}